\documentclass[sigconf]{acmart}

\usepackage{enumitem}
\usepackage{tabularx}
\usepackage{amsmath}
\usepackage{url}
\usepackage{balance}
\usepackage{multirow}
\begin{document}
\title{A Taxonomy and Dataset for 360$^\circ$ Videos}

\author{Afshin Taghavi Nasrabadi, Aliehsan Samiei, Anahita Mahzari, Ryan P. McMahan, Ravi Prakash}
\affiliation{%
  \institution{The University of Texas at Dallas, U.S.A.}
}
\email{{afshin, aliehsan.samiei, anahita.mahzari, rymcmaha, ravip}@utdallas.edu}

\author{Myl\`ene C.Q. Farias, Marcelo M. Carvalho}
\affiliation{
\institution{Department of Electrical Engineering \\ University of Brasilia (UnB)}}
\email{{mylene,mmcarvalho}@ene.unb.br}

\begin{abstract}
In this paper, we propose a taxonomy for $360^\circ$ videos that categorizes videos based on moving objects and camera motion. We gathered and produced 28 videos based on the taxonomy, and recorded viewport traces from 60 participants watching the videos. In addition to the viewport traces, we provide the viewers' feedback on their experience watching the videos, and we also analyze viewport patterns on each category.
\end{abstract}

\copyrightyear{2019} 
\acmYear{2019} 
\setcopyright{acmcopyright}
\acmConference[MMSys '19]{10th ACM Multimedia Systems Conference}{June 18--21, 2019}{Amherst, MA, USA}
\acmPrice{15.00}
\acmDOI{10.1145/3304109.3325812}
\acmISBN{978-1-4503-6297-9/19/06}


\maketitle

\section{Introduction}
Omnidirectional or $360^\circ$ video is one of the many Virtual Reality (VR) technologies with a growing popularity. $360^\circ$ video applications range from entertainment to education. These videos are usually watched through Head Mounted Displays (HMD) that enable viewers to explore a scene and look in any direction from a specific point in the scene. However, this new medium poses new challenges for content producers and service providers. For example, $360^\circ$ videos should have a high spatial resolution (4K or above) to provide an acceptable level of Quality of Experience (QoE) for viewers. Therefore, processing and streaming  this type of content is very demanding. 

Several solutions have been proposed to stream and render $360^\circ$ videos based on real-time users' viewport~\cite{corbillon2017viewport, nasrabadi2017adaptive}. They take advantage of the fact that users, at any point in time, view a limited portion of the video. To provide a high quality video inside a users' viewport, these methods need to know the users' viewport beforehand. This is typically done using viewport prediction methods. Since the accuracy and duration of existing viewport prediction methods are limited, viewport cannot be accurately predicted for time intervals longer than one second~\cite{nguyen2018your}. This limits the usefulness of viewport prediction under fluctuating network conditions as the video client has to buffer a long duration of the video to cope with network variations. Any mismatch between the predicted and actual user viewport can be detrimental to QoE. Another interesting challenge is storytelling, which, so far, has not been well-defined for this type of media. Unlike traditional videos, viewers are not limited to watch only a specific part of the scene determined by producers. Also, the effect of camera motion and scene-cuts are very different, which requires that producers know how to guide user attention. Currently, there are ongoing efforts to study the effect of different editing techniques on the QoE of $360^\circ$ videos~\cite{serrano2017movie}.

Solving the above-mentioned challenges requires study and analysis of user behaviors while watching $360^\circ$ videos. Publicly available viewport datasets facilitate these studies for several reasons. First, study of viewport traces enables content producers to understand which aspects of a $360^\circ$ video are important for users and how their attention can be guided. Second, datasets can help to develop and test viewport prediction and saliency detection methods. Moreover, these traces can be used by other researchers for the purpose of running experiments related to $360^\circ$ video streaming, as well as salience and visual attention modeling. Since users' viewport patterns are highly influenced by the video content, it is important to have various types of $360^\circ$ videos in a dataset. Moreover, a taxonomy of videos could be developed and videos could be appropriately classified on the basis of a set of attributes. If there is a high correlation between viewing patterns for videos in the same category, and significant differences between videos in different categories, then taxonomy information could be leveraged for viewport prediction. Today, there are several published datasets for users' viewport traces \cite{duanmu2018subjective}\cite{fremerey2018avtrack360}\cite{corbillon2017360}\cite{lo2017360}\cite{wu2017dataset}. However, they do not provide a comprehensive taxonomy of videos. 

We propose a taxonomy for $360^\circ$ videos that classifies videos based on camera motion and number of the moving targets in a video scene. We gathered and produced 28 scene-cut free videos based on the proposed taxonomy. We designed a subjective experiment in which 60 viewers watched a subset of the videos. In addition to providing the viewport traces for each viewing session, our dataset\footnote{ https://github.com/acmmmsys/2019-360dataset} includes the viewers' feedback about their experience after watching each video. Viewers described what they focused on and rated their perception of presence level of discomfort. These responses can be very helpful to study viewport traces. We also present preliminary analysis of user data that could be helpful in designing viewport prediction algorithms.

\section{Related Work}

Several datasets provide head movement traces of users watching $360^\circ$ videos. Corbillon \textit{et al.} \cite{corbillon2017360} gathered a dataset of viewport traces for five videos with 59 participants. Lo \textit{et al.} provided a dataset captured with 50 subjects watching 10 videos \cite{lo2017360}. Saliency and motion map of videos are also available in the dataset. The authors designed a viewport prediction method based on their dataset in \cite{fan2017fixation}. Wu \textit{et al.}~\cite{wu2017dataset} provided a dataset with 18 videos watched by 48 participants. They classified the videos based on their genre, such as sports, documentary, etc. This classification is very general and does not characterize the intrinsic properties of a scene. Their test procedure included two types of experiments. In the first experiment, subjects just watched the videos. In the second experiment, after each video, subjects were asked specific questions about the video content. This type of experiment forces viewers to pay attention to the content of videos. As a consequence, viewport samples are more scattered in the first experiment than in the second one. Duanmu \textit{et al.} \cite{duanmu2018subjective} provided a dataset of viewport traces in which videos were watched on a computer monitor. The authors compared the similarities and differences of viewing patterns between HMD and computer-based viewing sessions.

Fremerey \textit{et al.}~\cite{fremerey2018avtrack360} provided a dataset of head movements from 48 subjects watching twenty different videos. Participants also filled a Simulator Sickness (SS) Questionnaire after each set of five sequences. Although the overall discomfort was not very high, female participants experienced a higher discomfort level, which increased over the course of  different videos. David \textit{et al.} provided a dataset of head and eye movements~\cite{david2018dataset} for nineteen videos watched by 57 subjects. The dataset includes head+eye and head-only saliency maps and scan-paths. Interestingly, their results show that there are some differences between head-only and head+eye saliency maps, which are not highly correlated. According to the authors, this is caused by a loss of information in head-only maps. Generally, users' behavior depends on the content of the video. However, users' viewport is biased towards the center of the video. According to Fremerey \textit{et al.}~\cite{fremerey2018avtrack360}, most of the time users watch the areas closer to the center of the videos, with 90\% of the time within $\pm30^\circ$ deviation from equator, and 50\% of the time within the same interval from the center in horizontal direction. But, in horizontal direction, the viewports are more distributed, and nearly 50\% of users watched $330^\circ$ of horizontal view in all video sequences.

 In \cite{almquist2018prefetch}, thirty $360^\circ$ videos were shown to 32 participants. Videos were classified into 5 categories based on motion, but no distinction was made between the motion of object(s) in the scene versus camera motion. 
 Their analysis shows that viewport patterns change for different categories. Users viewport distribution along yaw axis tend to be more uniformly distributed for videos without moving objects. In another study \cite{ozcinar2018visual}, a dataset is created based on 6 videos with duration of 10 seconds, which is much shorter than previous studies. 17 participants have watched the videos, and each video was randomly repeated. They found that most of the fixation points are around moving objects. Moreover, videos with high motion complexity have fewer fixation points. Xu et al. \cite{xu2018predicting} have mined their own dataset for the purpose of viewport prediction. 58 subjects watched 76 panoramic videos. Analysis of their dataset shows that there is center bias, and there is similarity in the magnitude and direction of viewport changes when they are co-located.

\section{Taxonomy}
 Our goal is to design a taxonomy of $360^\circ$ videos that puts videos with similar user viewing patterns in the same category. Users' head movement can be triggered by different features of the content. Several studies on viewport dynamics suggest that user attention is guided by moving targets in the scene \cite{ozcinar2018visual}. Therefore, existence of moving objects plays an important role in a taxonomy. Additionally, we believe camera motion can affect viewer attention. In regular videos, camera motion dictates what users see in a scene. Although in $360^\circ$ videos users are free to look in any direction, camera motion can alter user behavior and transform the motion of moving objects. So, we also would like to study the effect of camera motion on user viewport.

Therefore, we propose a two dimensional taxonomy for $360^\circ$ videos based on the type of camera motion and number of moving objects. 
 We classify videos into five different categories based on camera motion: 1- Fixed, 2- Horizontal, 3- Vertical, 4- Rotational, 5- Mixture of previous camera motions. Regarding moving objects in a scene, we study the effect of the number of moving targets in a scene on viewport changes. So we have three categories: 1- No moving object, 2- Single moving object, 3- Multiple moving objects. By comparing videos from these categories, we can study the effect of moving objects on viewport pattern. This taxonomy results in a total of fifteen categories shown in Table \ref{tab:taxonomy} with each category corresponding to a $<$camera movement, number of moving targets$>$ combination.

\subsection{Videos used in the study}
\begin{figure}[htb]
\centering                            
    \centering
    \includegraphics[width=0.8\columnwidth]{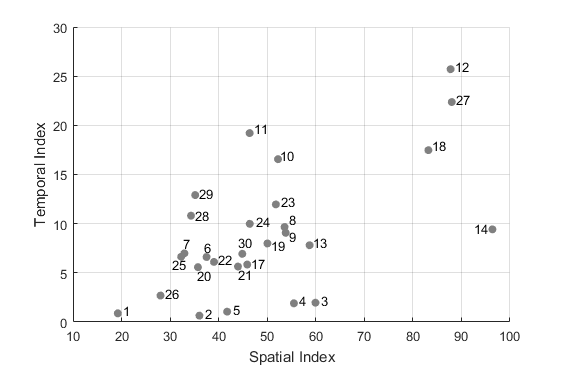}
    \caption{Spatial and Temporal complexity of dataset videos} \label{fig:siti}
\end{figure}

We considered two videos for each category in order to examine categorical similarities in viewport patterns. Each video has a duration of one minute with 
no scene cuts, so there is no discontinuity during each video. In Table~\ref{tab:taxonomy}, each video has a numerical ID from 1 to 30, referred to as $videoID$. Each cell contains two video IDs from the same category. For each video, resolution and frame rate are specified. Most of the videos were chosen from YouTube, but, we also produced several videos for categories such as rotational camera movements (videos 4,19,20,22,23,24). For the YouTube videos, links to the videos and the used start time, as many videos are longer than one minute, are inside brackets\footnote{Videos 10,17,27,28 were rotated 265, 180,63,81 degrees to right, respectively, to re-orient during playback.}. The full URL is the concatenation of www.youtu.be/ and the address in the table. For the recorded videos, we used a Samsung Gear360 camera to capture them. The spatial and temporal complexity \cite{ITUTP910} of the videos are depicted in Figure \ref{fig:siti}. In our study, because we focused on the visual stimuli only, we removed audio from the videos.

We did not include any video for the category of vertical camera motion with one moving object. 
To the best of our knowledge, there is no publicly available video containing only vertical camera motion and one moving object of at least one minute duration without scene cuts. We tried to produce two videos for this category using the Samsung Gear360 camera mounted on a drone, but the outcome was too shaky and could cause discomfort to viewers. So, we decided not to include the videos and category in our study.

\begin{table*}[htb]
\caption{Taxonomy and video links}
\resizebox{\textwidth}{!}{%
\begin{tabular}{ll|l|l|l|} 
\cline{3-5}
 &   &\multicolumn{3}{c|}{ \textbf{Number of Moving Targets}} \\\cline{3-5}
&   & \textbf{None}  & \textbf{Single} & \textbf{Multiple} \\ \hline
 \multicolumn{1}{|c|}  {\multirow{5}{*}{\rotatebox{90} {\textbf{ Camera Motion}}}} & \textbf{Fixed}      & \begin{tabular}[c]{@{}l@{}}1) 3840x1920 25fps [ESRz3-btZIA (0:40)] \\ 2) 3840x2160 29fps [30cSb3wTc7U (0:00)]\end{tabular} &   
\begin{tabular}[c]{@{}l@{}}3)  3840x2048 29fps [ULixPLH-WA4 (0:07)] \\ 4) 3840x1920 29fps
\end{tabular} &  
\begin{tabular}[c]{@{}l@{}}5)  3840x2048 30fps [7IWp875pCxQ (0:18)] \\ 6) 3840x2048 29fps [ze\_w7Lh97Co (0:05)]
\end{tabular}\\
\cline{2-5} 
\multicolumn{1}{|c|}{} & \textbf{Horizontal} &        
\begin{tabular}[c]{@{}l@{}}7)  3840x2160 25fps [9XR2CZi3V5k (0:01)]\\ 8)  3840x2048 29fps [6TlW1ClEBLY (0:45)]\end{tabular}&
\begin{tabular}[c]{@{}l@{}}9)  2560x1440 29fps [tVsw0DvAWdE (0:15)]\\ 10)  3840x1920 29fps [cNlQrTkXkOQ (0:15)]\end{tabular}& 
\begin{tabular}[c]{@{}l@{}}11)  3840x2160 24fps [jMyDqZe0z7M (0:00)]\\ 12)  3840x2160 30fps [2Lq86MKesG4 (0:12)]\end{tabular}\\ 
\cline{2-5} 
\multicolumn{1}{|c|}{} & \textbf{Vertical}   &
\begin{tabular}[c]{@{}l@{}}13)  3840x1920 29fps [DgxmQvWEGBU (0:04)]\\ 14)  3840x1920 29fps [elhdcvKhgbA (0:14)]\end{tabular}&
\begin{tabular}[c]{@{}l@{}}15) -----------------------    \\ 16) -----------------------   \end{tabular}&
\begin{tabular}[c]{@{}l@{}}17)  3840x2048 30fps [jau-Ric7kls (1:11)]\\ 18) 3840x2160 29fps [905\_oiaJN\_0 (0:15)]\end{tabular}\\ 
\cline{2-5} 
\multicolumn{1}{|c|}{} & \textbf{Rotational} & 
\begin{tabular}[c]{@{}l@{}}19) 3840x1920 29fps  \\ 20) 3840x1920 29fps\end{tabular}&
\begin{tabular}[c]{@{}l@{}}21)   3840x2160 29fps [ZRFIdczxxkY (0:04)] \\ 22) 3840x1920 29fps
\end{tabular}&
\begin{tabular}[c]{@{}l@{}}23) 3840x1920 29fps \\ 24) 3840x1920 29fps\end{tabular}\\ \cline{2-5} 
\multicolumn{1}{|c|}{}                               & \textbf{Mixed}      &
\begin{tabular}[c]{@{}l@{}}25)  3840x2048 25fps [HiRS\_6BCyG8 (0:26)]\\ 26) 3840x2160 29fps [L\_tqK4eqelA (5:30)]\end{tabular}&
\begin{tabular}[c]{@{}l@{}}27)  3840x1920 30fps [AX4hWfyHr5g (0:00)]\\ 28) 3840x1920 29fps [VGY4ksezNkY (2:11)]\end{tabular}&
\begin{tabular}[c]{@{}l@{}}29) 3840x2160 25fps [p9h3ZqJa1iA (0:00)]\\ 30) 3840x1906 29fps [H6SsB3JYqQg (1:00)]\end{tabular}\\ \hline
\end{tabular}
}
\label{tab:taxonomy}
\end{table*}

\section{Experimental methodology}
To study users' behavior under the proposed taxonomy, we designed a subjective experiment where participants watch a set of $360^\circ$ videos, using a HMD, and answer a set of questions after each video. The experiment has three main parts: 1) \textit{Training}, where participants answer an  entry questionnaire and watch an introductory video\footnote{https://youtu.be/mlOiXMvMaZo starting at 0:30}; 2) \textit{Main session}, where participants watch one video from each of the taxonomy categories and answer a questionnaire after each video; 3) \textit{Exit survey}, where participants answer a final questionnaire about their overall experience. Our experiment has been approved by the University Institutional Review Board.

The entry questionnaire is a background survey, which asks the participant's gender, age, and level of experience with using VR technology, including how many times the person has  watched $360^\circ$ videos. Then, the subject watches an introductory video to become familiar with the experiment and adjust the HMD. During the experiment, participants sit on a swivel chair and are free to look in any direction. We used Oculus Go HMD for this study, which is an all-in-one mobile, cable-free HMD. So, participants can rotate their head without cable interference. Since the videos do not have audio, users wear a headphone that plays white noise to eliminate auditory distractions. 

In the main session, each participant watches the videos in a shuffled order of categories to compensate for temporal effects. The shuffled list has one (of the two) video in each category. When the playback is finished for each video, a gray screen is shown. Then, the subject takes off the HMD and asked the following questions: 

\begin{itemize}
    \item[Q1)] Please describe what you saw while watching the video.
    
    \item[Q2)] What did you focus on while watching the video?
\end{itemize}

Then, participants rate their presence level. We used the following four questions from the self-location portion of the Spatial Presence Experience Scale \cite{hartmann2015spatial}:
\begin{itemize}
    \item[Q3)] I felt like I was actually there in the environment of the video.
    \item[Q4)] It seemed as though I actually took part in the action of the video. 
    \item[Q5)] It was as though my true location had shifted into the environment of the video.
    \item[Q6)] I felt as though I was physically present in the environment of the video.
\end{itemize}
Each question can be answered on a 5-point scale, from 1 (\textit{do not agree at all}) to 5 (\textit{fully agree}).

At the end of this questionnaire, we examine users' discomfort using a discomfort score~\cite{fernandes2016combating}. The participant chooses their discomfort score in the range from 0 to 10, where 10 is the highest discomfort level and 0 is the lowest. The experiment is terminated if a subject chooses the maximum score at any moment. At the end of the experiment, an exit survey is conducted that asks for participants to choose their three favorite videos. All questionnaires are answered on a desktop computer.

\subsection{Recording head movements}

We developed a video player in Unity using Pixvana SPINPlay SDK\footnote{https://pixvana.com/spin-sdk/} that plays back videos and records users' head orientation samples at the rate of HMD's screen refresh rate which is 60Hz. The player on the HMD is connected to a server on a PC. The server controls video playback on the HMD and collects recorded traces from HMD.

\begin{figure}[]
\centering                            
    \centering
    \includegraphics[width=\columnwidth]{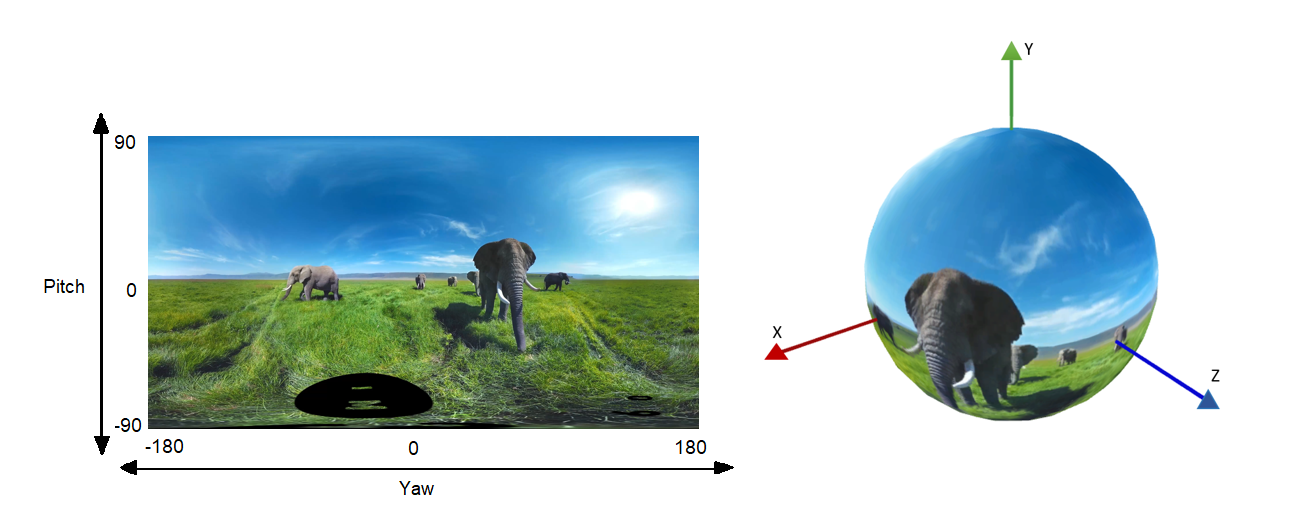}
    \caption{Left: Equirectangular frame. Right: The frame mapped to sphere and the coordinate system.} \label{fig:coord}
\end{figure}

Before presenting the format of the recorded viewport dataset, we explain the coordinate system and the video projection format  that were used. All videos are in equirectangular projection. Figure~\ref{fig:coord} depicts how an equirectangular video is shown to a viewer, along with the coordinate system. During playback, the video is mapped to a sphere. The coordinate system is defined such that the $Z$ axis always points out to the center of equirectangular video. If we assume that a viewer looks in the direction of $Z$, then the $Y$ axis points up and the $X$ axis points right. In the beginning of video playback, the sphere is rotated along the $Y$ axis to bring the center of the video to the front of the viewer. Note that the image on the surface of the sphere is mirrored, because the image is viewed from the inside of the sphere.

For each frame rendered on the HMD screen, we record the timestamp of the sample and the head rotation quaternion relative to the $Z$ axis. We use a quaternion coordinate system because it is able to represent a rotation with a higher accuracy, if compared to Euler angles. 
We get rotation samples from \texttt{UnityEngine.XR.InputTracking} class in Unity. \texttt{GetLocalRotation(XRNode.CenterEye)} function provides the quaternion rotation of VR HMD. We record the samples and playback timestamp (with millisecond accuracy) inside \texttt{Update()} function loop that runs per rendered frame. In addition to the quaternion rotation, we also include the Cartesian form of the vector that points out to the center of user's viewport. Therefore, the format of the recorded samples is the following:    $$\{timestamp,Q_x,Q_y,Q_z,Q_w,V_x,V_y,V_z\}$$
where $\{Q_x,Q_y,Q_z,Q_w\}$ denotes the components of the quaternion of viewport rotation and $\{V_x,V_y,V_z\}$ specifies the vector from the sphere center to viewport center. We store all samples in a CSV file.

\subsection{Dataset Structure}
Our dataset has six main data folders: \textit{Traces}, \textit{Questionnaires}, \textit{ViewportHeatmap}, \textit{SampleVideos}, \textit{Scripts}, and \textit{Histograms}.  The \textit{Traces} folder contains the viewport traces of all participants. Each participant is assigned a 6-character number: the $SubjectID$. The \textit{Traces} folder contains one sub-folder for each participant, named according to this $SubjectID$. It contains a CSV file, i.e., a trace file for each video watched by this participant. The CSV filename format is $SubjectID\_VideoID.csv$. The \textit{Questionnaires} folder contains the participants' responses to all questions. The \textit{BackgroundQuestionnaire.csv} contains responses to the background survey and the \textit{PerVideoQuestionnaire.csv} contains the answers to the questionnaire completed after each video. The header of questionnaire files contains the questions.

We created a heatmap of viewport traces for each video. All heatmaps are included in the \textit{ViewportHeatmap} folder. We represent each subject's viewport center using a $10^\circ$ Gaussian kernel and created a heatmap for each frame by applying the kernel for all thirty viewers of a video. The \textit{SampleVideos} folder contains a subset of videos in the taxonomy\footnote{Only the videos that we have created.}. The video sources are in the \textit{Source} folder, and viewport-overlaid versions are in the \textit{ViewportOverlays} folder. Heatmap of each video was merged with its original video to create each viewport-overlaid video\footnote{for all videos in the taxonomy, the overlaid version can be found in: https://bit.ly/2P1DRR7.}. These videos provide useful visual representations of how the videos were watched by viewers. Figure \ref{fig:ab} shows example frames of a viewport-overlaid video. The \textit{scripts} folder includes MATLAB scripts to generate overlay videos from viewport traces and source of videos. The codes for running clustering on viewport traces are also included in the \textit{Scripts} folder. Finally, the \textit{Histogram} folder contains histograms of yaw and pitch angles for each video.

\section{Results}
A total of 60 persons participated in our study, from which 28.3\% of participants were female. Each subject watched 14 videos plus the introductory video, with each video being watched by 30 viewers.  Table~\ref{tab:summary} shows the age distribution of experiment participants, along with their experience with VR technologies. 

\begin{table}[htb]
\caption{Subjects distribution and VR experience}
\resizebox{\columnwidth}{!}{%
\begin{tabular}{|l|l|l|l|l|}
\hline
\textbf{Gender} & \textbf{Age}  & \textbf{Mobile VR Exp.}  & \textbf{Room Scale VR Exp.}  & \textbf{360 Exp.} \\ \hline
\begin{tabular}[c]{@{}l@{}} 17 Female\\ 43 Male\end{tabular} & \begin{tabular}[c]{@{}l@{}}18-21 : 19\\ 22-25 : 16\\ 26-29 : 15\\ 30-33 : 7\\ \textgreater{}40 : 3\end{tabular} & \begin{tabular}[c]{@{}l@{}}Never: 17\\ 1-5 times: 31\\ 6-10 times: 4\\ 11-20 times: 3\\ \textgreater{}20 times: 5\end{tabular} & \begin{tabular}[c]{@{}l@{}}Never: 37\\ 1-5 times: 15\\ 6-10 times: 5\\ \textgreater{}10 times: 3\end{tabular} & \begin{tabular}[c]{@{}l@{}}Never: 21\\ 1-5 videos: 28\\ 6-10 videos: 6\\ \textgreater{}10 videos: 5\end{tabular}  \\ \hline
\end{tabular}
}
\label{tab:summary}
\end{table}

\subsection{Viewport Pattern}

Analysis of the distributions of yaw and pitch angles over the whole duration of videos shows that pitch angle is biased along the equatorial section of the video. However, for videos captured from higher altitudes, the samples are more distributed, e.g., videos 17, 18, and 27. For the yaw angle, the distribution depends on the content and location of regions of interest. For example, videos 23 and 24, which contain rotational camera motion with multiple moving objects, have an almost uniform angle distribution (see `histogram' folder).

\begin{figure}[htb]
\centering                            
    \centering
    \includegraphics[width=0.9\columnwidth]{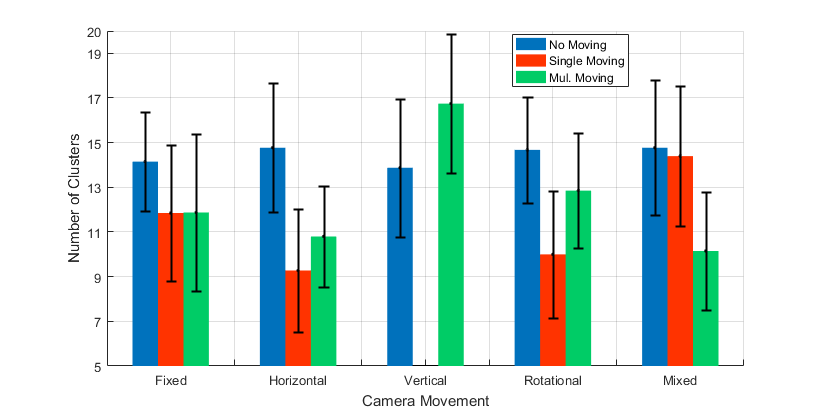}
    \caption{Barplot of the average number of clusters per category, along with the error bars.} \label{fig:groupCluster}
\end{figure}

\begin{figure}[htb]
\centering                            
    \centering
    \includegraphics[width=0.85\columnwidth]{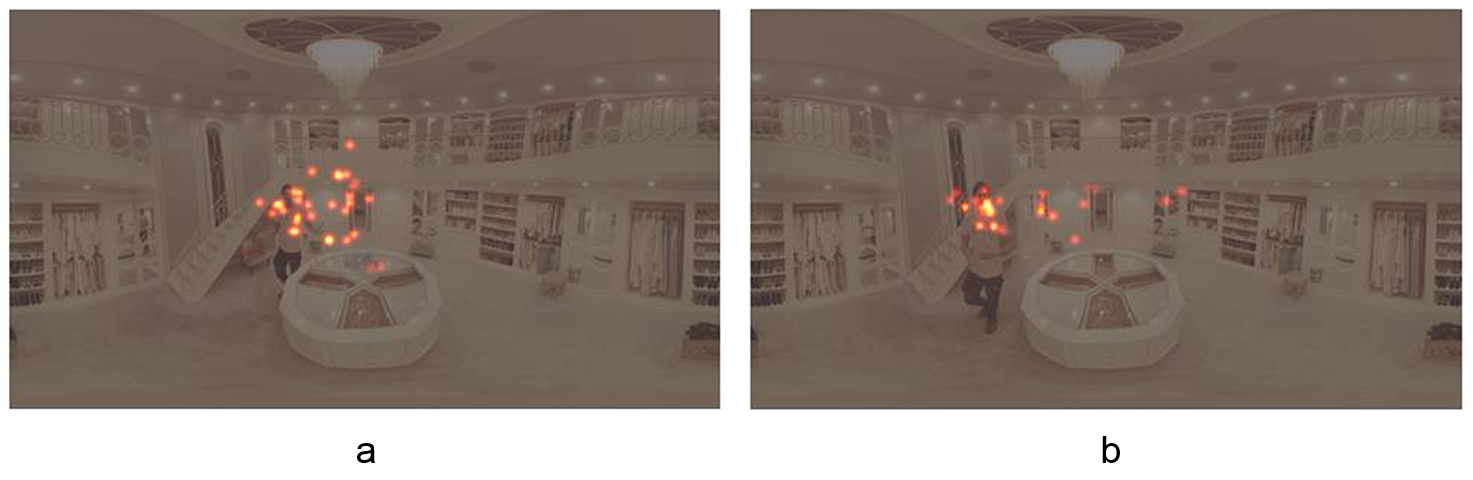}
    \caption{Two frames from video 3 that show  viewers tracking a moving target (walking man).} \label{fig:ab}
\end{figure}

To analyze users' viewport pattern, we use the clustering algorithm proposed by Rossi \textit{et al.} \cite{rossi2018spherical} which clusters viewports based on their overlap in spherical domain. We use an angle threshold of $\pi/5$. The clustering algorithm divides a video into 3-second chunks, and viewports which are less than $\pi/5$ apart for 60\% of the duration of a chunk are placed in the same cluster. A small number of clusters means that  most viewers focused on specific parts of the video, while a high number of them means that there were no common central focal points. Figure~\ref{fig:groupCluster} shows the average number of viewport clusters for each category, with the error bars indicating the standard deviation associated to the averages. Notice that the number of clusters decreases as moving targets are added to the scene, with the exception being the videos with  vertical camera movement.  Figure \ref{fig:ab} shows how a large number of viewers tracked the moving targets from frame \textit{a} to frame \textit{b}, which is one second later. Notice from this figure that a group of observers seem to be following moving targets (in this case, the man in the frames).

For the category corresponding to vertical camera movement and multiple moving targets, viewports are more dispersed compared to no moving target videos (see green bar in the third group in Figure~\ref{fig:groupCluster}). One possible  reason for this behavior is that for most of the duration of these videos, the camera is located at high distance from the ground level and viewers have a landscape view. The viewport-overlaid version of these videos show that viewers were more interested in the landscape view, and did not focus on any specific area of the video.

\begin{figure}[htb]
\centering                            
    \centering
    \includegraphics[width=\columnwidth]{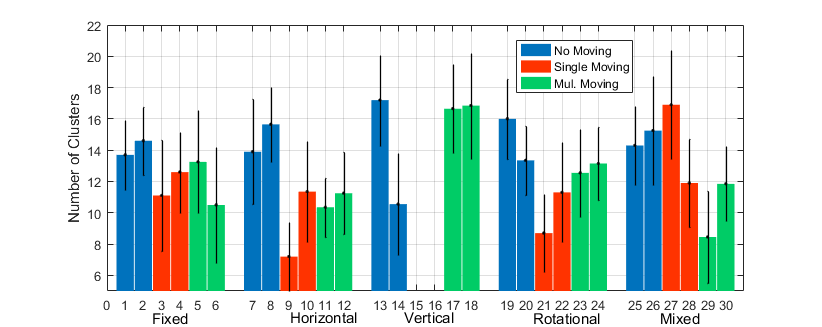}
    \caption{Barplot of the average number of clusters per video, along with the corresponding error bars.} \label{fig:videoCluster}
\end{figure}

\begin{figure}[htb]
\centering                            
    \centering
    \includegraphics[width=\columnwidth]{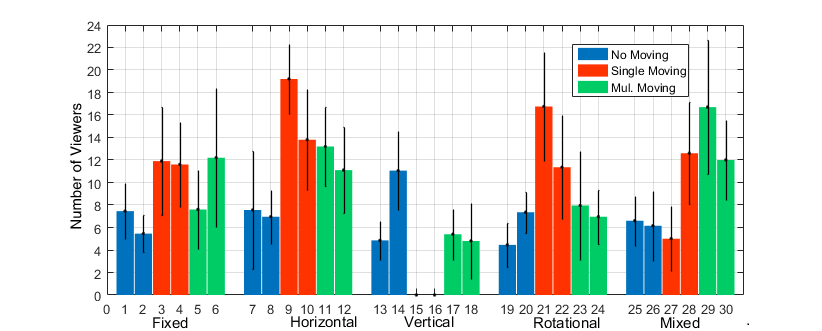}
    \caption{Barplot of the average number of viewers per video in the most populated cluster, along with  the error bars.} \label{fig:pervideoMaxinCluster}
\end{figure}

Figure \ref{fig:videoCluster} shows the clustering results separated per video (and ordered by category), and Figure \ref{fig:pervideoMaxinCluster} depicts the corresponding number of viewers in the largest cluster for each of these videos. Notice that in some categorical pairs, such as (9,10), (13,14), and (27,28), the number of clusters for the videos is very different. For example,  video 9 shows a view from a racing car that chases another car, while video 10 shows a woman walking in the woods. Both these videos were classified as having a horizontal movement and a single moving target. But, although for both videos the viewport-overlaid versions show a high concentration of viewports on the moving target, the speed of camera for video 9 is much higher and there are few objects (other than the two cars) in the video. Most likely, for single moving target videos, when the camera movement direction is aligned with the moving target, viewers are influenced to look at the target. The number of clusters is smaller for this type of videos.

\begin{figure}[]
\centering                            
    \centering
    \includegraphics[width=0.8\columnwidth]{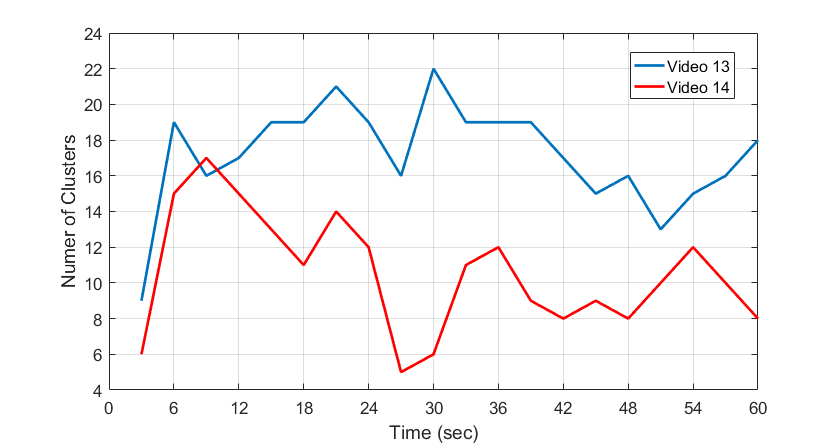}
    \caption{Number of clusters for videos 13 and 14} \label{fig:video1314}
\end{figure}

\begin{figure}[]
\centering                            
    \centering
    \includegraphics[width=0.9\columnwidth]{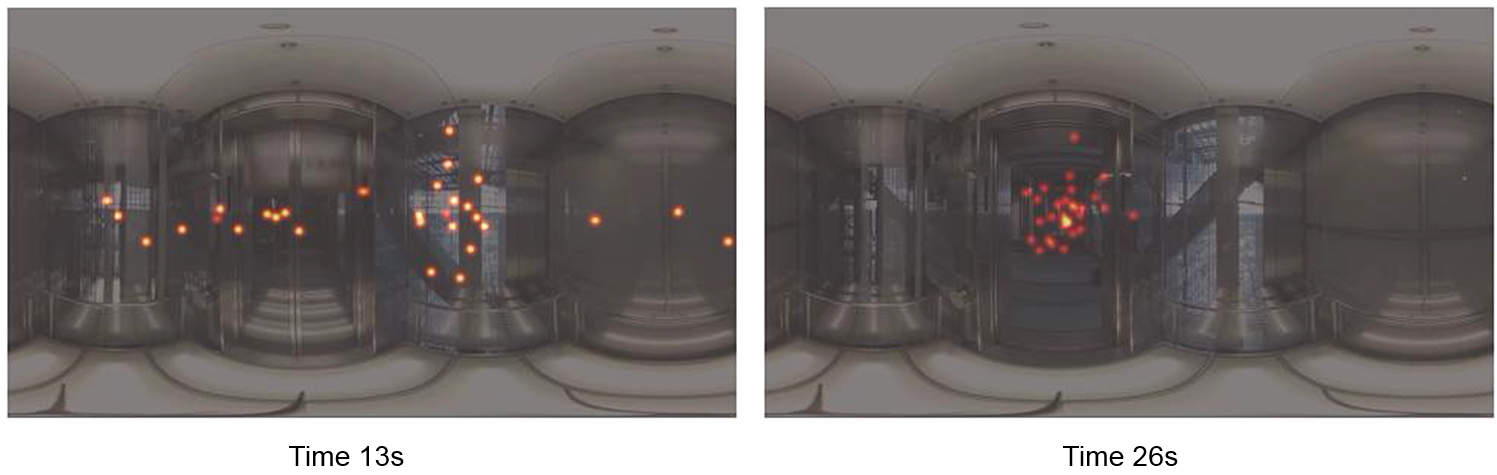}
    \caption{Two viewport-overlaid frames of video 14} \label{fig:video14}
\end{figure}

Videos 13 and 14 have a view from inside a glass elevator, with a vertical camera motion and no moving targets. One of the main differences between these two videos is that in video 14 (at time 0:22) the elevator stops and the door opens, attracting a lot of the viewers' attention and, therefore, reducing the number of clusters. Figure \ref{fig:video1314} shows how the number of clusters changes over time for videos 13 and 14, where a drop in the number of clusters can be seen in the curve for video 14 after 22s. 
Looking more closely, Figure \ref{fig:video14} shows the viewport-overlaid views of the frames at instant 14s and 26s of video 14. Videos 17, 18, and 27 are also shot from a high altitude, and based on viewers feedback and viewport-overlaid videos, also had a landscape view that viewers found more interesting. 

Comparing categories with one moving target, videos with camera motion can have fewer clusters if the camera follows the target. For example, in videos 8, 9, and 21, the camera moves according to the moving target, and the number of clusters is less compared to fixed camera category. In video 21, moving target is always at the center of video, and this video has more viewers in one cluster compared to similar videos without camera motion, e.g., videos 3 and 4. Video 22 has camera rotation but it does not follow the moving target, and viewers were more dispersed compared to video 21. However, for mixed camera motion, the pattern is not the same. The video scenery and camera elevation affects the viewers.

\begin{figure}[]
\centering                            
    \centering
    \includegraphics[width=\columnwidth]{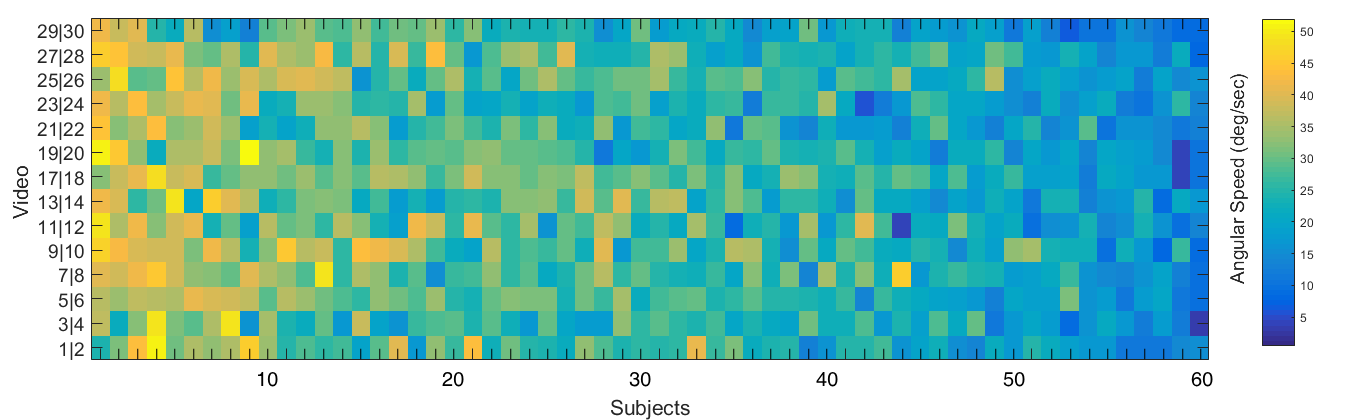}
    \caption{Subjects' head-movement speed heatmap} \label{fig:avgstd_sorted}
\end{figure}

We also observed that viewers have different head movement and angular speed patterns. Some viewers tend to be still and rotate their head occasionally compared to others. Figure~\ref{fig:avgstd_sorted} shows the heatmap of average angular speed of viewers for each video. We  measured the  head movement speed at  time intervals of 1 second, which is the great-circle distance between two head orientation samples divided by the total time. In this figure, each row corresponds to a category in the taxonomy and each column to a viewer. Viewers are sorted according to their average angular speed over all videos. It can be observed that viewers with a higher average speed watched all videos with a higher speed, which are represented by bright yellow colors in the map in Figure~\ref{fig:avgstd_sorted}. On the other hand, barring a few exceptions, slower viewers had smaller average speeds, which are represented by darker blue colors in the map, for all videos. This suggests that users could be possibly classified on the basis of their head movement speed. So, head movement speed of a specific user in previous viewing sessions could be used to predict his/her future viewport.

\subsection{Questionnaires}
\begin{figure}[]
\centering                            
    \centering
    \includegraphics[width=\columnwidth]{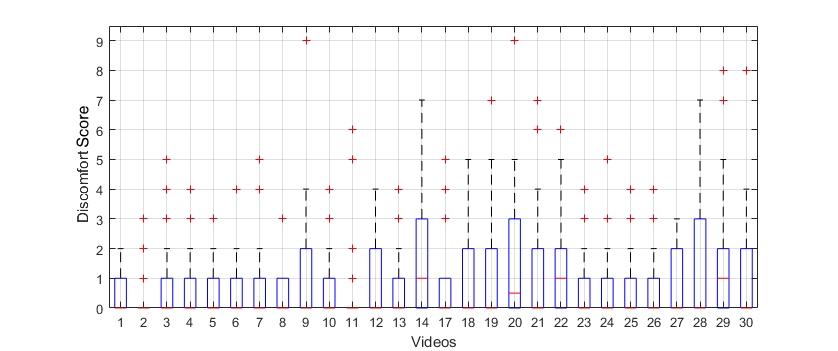}
    \caption{Boxplot of discomfort score per video} \label{fig:discomfort}
\end{figure}

After watching each video, viewers chose their discomfort level. Figure \ref{fig:discomfort} shows the box-plot graph of the users' responses. Although the range of scores is from 0 to 10, for most videos, the median score value was close to 0. More specifically, videos with fixed camera movements received low discomfort scores, while  videos with camera motion received slightly higher discomfort scores. Also, for the presence level questions, Q3 to Q6, the average score is around 3, and the detailed responses are available in the dataset.

\section{Conclusion}
In this paper, we presented a taxonomy and dataset for $360^\circ$ videos. We analyzed viewport traces according to the number of viewer clusters in each video. Generally videos with moving targets have fewer clusters, and preliminary investigation of viewport overlays on videos suggest that users tend to look at moving targets. However, there are some exceptions based on camera location and video scenery. For example, we observed that videos captured from higher altitudes have more dispersed viewport distribution irrespective of the number of moving objects. Some viewers tend to explore the scene more aggressively while others tend to be more passive, regardless of the nature of the video. In future work, we will study the effect of moving objects on viewport pattern in more detail.

\bibliographystyle{ACM-Reference-Format}
\balance
\bibliography{main} 


\begin{thebibliography}{00}


\ifx \showCODEN    \undefined \def \showCODEN     #1{\unskip}     \fi
\ifx \showDOI      \undefined \def \showDOI       #1{{\tt DOI:}\penalty0{#1}\ }
  \fi
\ifx \showISBNx    \undefined \def \showISBNx     #1{\unskip}     \fi
\ifx \showISBNxiii \undefined \def \showISBNxiii  #1{\unskip}     \fi
\ifx \showISSN     \undefined \def \showISSN      #1{\unskip}     \fi
\ifx \showLCCN     \undefined \def \showLCCN      #1{\unskip}     \fi
\ifx \shownote     \undefined \def \shownote      #1{#1}          \fi
\ifx \showarticletitle \undefined \def \showarticletitle #1{#1}   \fi
\ifx \showURL      \undefined \def \showURL       #1{#1}          \fi
\providecommand\bibfield[2]{#2}
\providecommand\bibinfo[2]{#2}
\providecommand\natexlab[1]{#1}
\providecommand\showeprint[2][]{arXiv:#2}

\bibitem[\protect\citeauthoryear{Almquist, Almquist, Krishnamoorthi, Carlsson,
  and Eager}{Almquist et~al\mbox{.}}{2018}]%
        {almquist2018prefetch}
\bibfield{author}{\bibinfo{person}{Mathias Almquist}, \bibinfo{person}{Viktor
  Almquist}, \bibinfo{person}{Vengatanathan Krishnamoorthi},
  \bibinfo{person}{Niklas Carlsson}, {and} \bibinfo{person}{Derek Eager}.}
  \bibinfo{year}{2018}\natexlab{}.
\newblock \showarticletitle{The Prefetch Aggressiveness Tradeof in 360 Video
  Streaming}. In \bibinfo{booktitle}{{\em Proceedings of ACM Multimedia Systems
  Conference. Amsterdam, Netherlands}}.
\newblock


\bibitem[\protect\citeauthoryear{Corbillon, De~Simone, and Simon}{Corbillon
  et~al\mbox{.}}{2017a}]%
        {corbillon2017360}
\bibfield{author}{\bibinfo{person}{Xavier Corbillon},
  \bibinfo{person}{Francesca De~Simone}, {and} \bibinfo{person}{Gwendal
  Simon}.} \bibinfo{year}{2017}\natexlab{a}.
\newblock \showarticletitle{360-degree video head movement dataset}. In
  \bibinfo{booktitle}{{\em Proceedings of the 8th ACM on Multimedia Systems
  Conference}}. ACM, \bibinfo{pages}{199--204}.
\newblock


\bibitem[\protect\citeauthoryear{Corbillon, Simon, Devlic, and
  Chakareski}{Corbillon et~al\mbox{.}}{2017b}]%
        {corbillon2017viewport}
\bibfield{author}{\bibinfo{person}{Xavier Corbillon}, \bibinfo{person}{Gwendal
  Simon}, \bibinfo{person}{Alisa Devlic}, {and} \bibinfo{person}{Jacob
  Chakareski}.} \bibinfo{year}{2017}\natexlab{b}.
\newblock \showarticletitle{Viewport-adaptive navigable 360-degree video
  delivery}. In \bibinfo{booktitle}{{\em Communications (ICC), 2017 IEEE
  International Conference on}}. IEEE, \bibinfo{pages}{1--7}.
\newblock


\bibitem[\protect\citeauthoryear{David, Guti{\'e}rrez, Coutrot, Da~Silva, and
  Callet}{David et~al\mbox{.}}{2018}]%
        {david2018dataset}
\bibfield{author}{\bibinfo{person}{Erwan~J David}, \bibinfo{person}{Jes{\'u}s
  Guti{\'e}rrez}, \bibinfo{person}{Antoine Coutrot},
  \bibinfo{person}{Matthieu~Perreira Da~Silva}, {and}
  \bibinfo{person}{Patrick~Le Callet}.} \bibinfo{year}{2018}\natexlab{}.
\newblock \showarticletitle{A dataset of head and eye movements for 360$^\circ$
  videos}. In \bibinfo{booktitle}{{\em Proceedings of the 9th ACM Multimedia
  Systems Conference}}. ACM, \bibinfo{pages}{432--437}.
\newblock


\bibitem[\protect\citeauthoryear{Duanmu, Mao, Liu, Srinivasan, and Wang}{Duanmu
  et~al\mbox{.}}{2018}]%
        {duanmu2018subjective}
\bibfield{author}{\bibinfo{person}{Fanyi Duanmu}, \bibinfo{person}{Yixiang
  Mao}, \bibinfo{person}{Shuai Liu}, \bibinfo{person}{Sumanth Srinivasan},
  {and} \bibinfo{person}{Yao Wang}.} \bibinfo{year}{2018}\natexlab{}.
\newblock \showarticletitle{A Subjective Study of Viewer Navigation Behaviors
  When Watching 360-Degree Videos on Computers}. In \bibinfo{booktitle}{{\em
  2018 IEEE International Conference on Multimedia and Expo (ICME)}}. IEEE,
  \bibinfo{pages}{1--6}.
\newblock


\bibitem[\protect\citeauthoryear{Fan, Lee, Lo, Huang, Chen, and Hsu}{Fan
  et~al\mbox{.}}{2017}]%
        {fan2017fixation}
\bibfield{author}{\bibinfo{person}{Ching-Ling Fan}, \bibinfo{person}{Jean Lee},
  \bibinfo{person}{Wen-Chih Lo}, \bibinfo{person}{Chun-Ying Huang},
  \bibinfo{person}{Kuan-Ta Chen}, {and} \bibinfo{person}{Cheng-Hsin Hsu}.}
  \bibinfo{year}{2017}\natexlab{}.
\newblock \showarticletitle{Fixation prediction for 360 video streaming in
  head-mounted virtual reality}. In \bibinfo{booktitle}{{\em Proceedings of the
  27th Workshop on Network and Operating Systems Support for Digital Audio and
  Video}}. ACM, \bibinfo{pages}{67--72}.
\newblock


\bibitem[\protect\citeauthoryear{Fernandes and Feiner}{Fernandes and
  Feiner}{2016}]%
        {fernandes2016combating}
\bibfield{author}{\bibinfo{person}{Ajoy~S Fernandes} {and}
  \bibinfo{person}{Steven~K Feiner}.} \bibinfo{year}{2016}\natexlab{}.
\newblock \showarticletitle{Combating VR sickness through subtle dynamic
  field-of-view modification}. In \bibinfo{booktitle}{{\em 3D User Interfaces
  (3DUI), 2016 IEEE Symposium on}}. IEEE, \bibinfo{pages}{201--210}.
\newblock


\bibitem[\protect\citeauthoryear{Fremerey, Singla, Meseberg, and
  Raake}{Fremerey et~al\mbox{.}}{2018}]%
        {fremerey2018avtrack360}
\bibfield{author}{\bibinfo{person}{Stephan Fremerey}, \bibinfo{person}{Ashutosh
  Singla}, \bibinfo{person}{Kay Meseberg}, {and} \bibinfo{person}{Alexander
  Raake}.} \bibinfo{year}{2018}\natexlab{}.
\newblock \showarticletitle{AVtrack360: an open dataset and software recording
  people's head rotations watching 360° videos on an HMD}. In
  \bibinfo{booktitle}{{\em Proceedings of the 9th ACM Multimedia Systems
  Conference}}. ACM, \bibinfo{pages}{403--408}.
\newblock


\bibitem[\protect\citeauthoryear{Hartmann, Wirth, Schramm, Klimmt, Vorderer,
  Gysbers, B{\"o}cking, Ravaja, Laarni, Saari, et~al\mbox{.}}{Hartmann
  et~al\mbox{.}}{2015}]%
        {hartmann2015spatial}
\bibfield{author}{\bibinfo{person}{Tilo Hartmann}, \bibinfo{person}{Werner
  Wirth}, \bibinfo{person}{Holger Schramm}, \bibinfo{person}{Christoph Klimmt},
  \bibinfo{person}{Peter Vorderer}, \bibinfo{person}{Andr{\'e} Gysbers},
  \bibinfo{person}{Saskia B{\"o}cking}, \bibinfo{person}{Niklas Ravaja},
  \bibinfo{person}{Jari Laarni}, \bibinfo{person}{Timo Saari}, {and}
  \bibinfo{person}{others}.} \bibinfo{year}{2015}\natexlab{}.
\newblock \showarticletitle{The spatial presence experience scale (SPES)}.
\newblock \bibinfo{journal}{{\em Journal of Media Psychology\/}}
  (\bibinfo{year}{2015}).
\newblock


\bibitem[\protect\citeauthoryear{{ITU-T Recommendation P.910}}{{ITU-T
  Recommendation P.910}}{2008}]%
        {ITUTP910}
\bibfield{author}{\bibinfo{person}{{ITU-T Recommendation P.910}}.}
  \bibinfo{year}{2008}\natexlab{}.
\newblock \bibinfo{title}{{Subjective video quality assessment methods for
  multimedia applications}}.
\newblock   (\bibinfo{date}{April} \bibinfo{year}{2008}).
\newblock


\bibitem[\protect\citeauthoryear{Lo, Fan, Lee, Huang, Chen, and Hsu}{Lo
  et~al\mbox{.}}{2017}]%
        {lo2017360}
\bibfield{author}{\bibinfo{person}{Wen-Chih Lo}, \bibinfo{person}{Ching-Ling
  Fan}, \bibinfo{person}{Jean Lee}, \bibinfo{person}{Chun-Ying Huang},
  \bibinfo{person}{Kuan-Ta Chen}, {and} \bibinfo{person}{Cheng-Hsin Hsu}.}
  \bibinfo{year}{2017}\natexlab{}.
\newblock \showarticletitle{360 video viewing dataset in head-mounted virtual
  reality}. In \bibinfo{booktitle}{{\em Proceedings of the 8th ACM on
  Multimedia Systems Conference}}. ACM, \bibinfo{pages}{211--216}.
\newblock


\bibitem[\protect\citeauthoryear{Nasrabadi, Mahzari, Beshay, and
  Prakash}{Nasrabadi et~al\mbox{.}}{2017}]%
        {nasrabadi2017adaptive}
\bibfield{author}{\bibinfo{person}{Afshin~Taghavi Nasrabadi},
  \bibinfo{person}{Anahita Mahzari}, \bibinfo{person}{Joseph~D Beshay}, {and}
  \bibinfo{person}{Ravi Prakash}.} \bibinfo{year}{2017}\natexlab{}.
\newblock \showarticletitle{Adaptive 360-degree video streaming using scalable
  video coding}. In \bibinfo{booktitle}{{\em Proceedings of the 2017 ACM on
  Multimedia Conference}}. ACM, \bibinfo{pages}{1689--1697}.
\newblock


\bibitem[\protect\citeauthoryear{Nguyen, Yan, and Nahrstedt}{Nguyen
  et~al\mbox{.}}{2018}]%
        {nguyen2018your}
\bibfield{author}{\bibinfo{person}{Anh Nguyen}, \bibinfo{person}{Zhisheng Yan},
  {and} \bibinfo{person}{Klara Nahrstedt}.} \bibinfo{year}{2018}\natexlab{}.
\newblock \showarticletitle{Your Attention is Unique: Detecting 360-Degree
  Video Saliency in Head-Mounted Display for Head Movement Prediction}. In
  \bibinfo{booktitle}{{\em 2018 ACM Multimedia Conference on Multimedia
  Conference}}. ACM, \bibinfo{pages}{1190--1198}.
\newblock


\bibitem[\protect\citeauthoryear{Ozcinar and Smolic}{Ozcinar and
  Smolic}{2018}]%
        {ozcinar2018visual}
\bibfield{author}{\bibinfo{person}{Cagri Ozcinar} {and} \bibinfo{person}{Aljosa
  Smolic}.} \bibinfo{year}{2018}\natexlab{}.
\newblock \showarticletitle{Visual attention in omnidirectional video for
  virtual reality applications}. In \bibinfo{booktitle}{{\em 2018 Tenth
  International Conference on Quality of Multimedia Experience (QoMEX)}}. IEEE,
  \bibinfo{pages}{1--6}.
\newblock


\bibitem[\protect\citeauthoryear{Rossi, De~Simone, Frossard, and Toni}{Rossi
  et~al\mbox{.}}{2018}]%
        {rossi2018spherical}
\bibfield{author}{\bibinfo{person}{Silvia Rossi}, \bibinfo{person}{Francesca
  De~Simone}, \bibinfo{person}{Pascal Frossard}, {and} \bibinfo{person}{Laura
  Toni}.} \bibinfo{year}{2018}\natexlab{}.
\newblock \showarticletitle{Spherical clustering of users navigating
  360$^\circ$ content}.
\newblock \bibinfo{journal}{{\em arXiv preprint arXiv:1811.05185\/}}
  (\bibinfo{year}{2018}).
\newblock


\bibitem[\protect\citeauthoryear{Serrano, Sitzmann, Ruiz-Borau, Wetzstein,
  Gutierrez, and Masia}{Serrano et~al\mbox{.}}{2017}]%
        {serrano2017movie}
\bibfield{author}{\bibinfo{person}{Ana Serrano}, \bibinfo{person}{Vincent
  Sitzmann}, \bibinfo{person}{Jaime Ruiz-Borau}, \bibinfo{person}{Gordon
  Wetzstein}, \bibinfo{person}{Diego Gutierrez}, {and} \bibinfo{person}{Belen
  Masia}.} \bibinfo{year}{2017}\natexlab{}.
\newblock \showarticletitle{Movie editing and cognitive event segmentation in
  virtual reality video}.
\newblock \bibinfo{journal}{{\em ACM Transactions on Graphics (TOG)\/}}
  \bibinfo{volume}{36}, \bibinfo{number}{4} (\bibinfo{year}{2017}),
  \bibinfo{pages}{47}.
\newblock


\bibitem[\protect\citeauthoryear{Wu, Tan, Wang, and Yang}{Wu
  et~al\mbox{.}}{2017}]%
        {wu2017dataset}
\bibfield{author}{\bibinfo{person}{Chenglei Wu}, \bibinfo{person}{Zhihao Tan},
  \bibinfo{person}{Zhi Wang}, {and} \bibinfo{person}{Shiqiang Yang}.}
  \bibinfo{year}{2017}\natexlab{}.
\newblock \showarticletitle{A Dataset for Exploring User Behaviors in VR
  Spherical Video Streaming}. In \bibinfo{booktitle}{{\em Proceedings of the
  8th ACM on Multimedia Systems Conference}}. ACM, \bibinfo{pages}{193--198}.
\newblock


\bibitem[\protect\citeauthoryear{Xu, Song, Wang, Qiao, Huo, and Wang}{Xu
  et~al\mbox{.}}{2018}]%
        {xu2018predicting}
\bibfield{author}{\bibinfo{person}{Mai Xu}, \bibinfo{person}{Yuhang Song},
  \bibinfo{person}{Jianyi Wang}, \bibinfo{person}{MingLang Qiao},
  \bibinfo{person}{Liangyu Huo}, {and} \bibinfo{person}{Zulin Wang}.}
  \bibinfo{year}{2018}\natexlab{}.
\newblock \showarticletitle{Predicting head movement in panoramic video: A deep
  reinforcement learning approach}.
\newblock \bibinfo{journal}{{\em IEEE transactions on pattern analysis and
  machine intelligence\/}} (\bibinfo{year}{2018}).
\newblock


\end{thebibliography}

\end{document}